\documentclass[aps,pra,twocolumn,superscriptaddress,nofootinbib,amsmath,amssymb]{revtex4-2}

\usepackage{dcolumn}                    
\usepackage{bm}                         
\usepackage[pdftex]{graphicx,color}     
\usepackage{graphicx}                   
\usepackage{float}                      
\usepackage{svg}                        
\usepackage[english]{babel}             
\usepackage[normalem]{ulem}             

\begin{document}

\title{Observation of mutual extinction and transparency in light scattering}

\author{Alfredo Rates}
\affiliation{Complex Photonic Systems (COPS), MESA+ Institute for Nanotechnology, 
University of Twente, P.O. Box 217, 7500 AE Enschede, The Netherlands}

\author{Ad Lagendijk}
\email{a.lagendijk@utwente.nl}
\affiliation{Complex Photonic Systems (COPS), MESA+ Institute for Nanotechnology, 
University of Twente, P.O. Box 217, 7500 AE Enschede, The Netherlands}

\author{Ozan Akdemir}
\affiliation{Complex Photonic Systems (COPS), MESA+ Institute for Nanotechnology, 
University of Twente, P.O. Box 217, 7500 AE Enschede, The Netherlands}

\author{Allard P. Mosk}
\affiliation{Debye Institute for NanoMaterials Science and Center for Extreme Matter and Emergent Phenomena, 
Utrecht University, Princetonplein 5, 3584 CC Utrecht, The Netherlands}

\author{Willem L. Vos}
\email{w.l.vos@utwente.nl}
\affiliation{Complex Photonic Systems (COPS), MESA+ Institute for Nanotechnology, 
University of Twente, P.O. Box 217, 7500 AE Enschede, The Netherlands}

\date{August 13th, 2021.}


\begin{abstract}
Interference of scattered waves is fundamental for modern light-scattering techniques, such as optical wavefront shaping. 
Recently, a new type of wavefront shaping was introduced where the extinction is manipulated instead of the scattered intensity. 
The underlying idea is that upon changing the phases or the amplitudes of incident beams, the total extinction will change due to interference described by the cross terms between different incident beams. 
Here, we experimentally demonstrate the mutual extinction and transparency effects in scattering media, in particular, a human hair and a silicon bar. 
To this end, we send two light beams with a variable mutual angle on the sample. 
Depending on the relative phase of the incident beams we observe either nearly zero extinction, mutual transparency, or almost twice the single-beam extinction, mutual extinction, in agreement with theory. 
We use an analytical approximation for the scattering amplitude, starting from a completely opaque object and we discuss the limitations of our approximation. 
We discuss the applications of the mutual extinction and transparency effects in various fields such as non-line-of-sight communications, microscopy, and biomedical imaging. 
\end{abstract}
\maketitle

%
\section{Introduction}
%
Random scattering of light inside complex materials such as clouds, paint, milk, white LEDs, hair, or human tissue is what makes them appear opaque~\cite{ishimaru1978book,VanRossum1999rmp,Schubert2006book,Wiersma2013np,Meretska2019acs}. 
In these inhomogeneous materials, multiple scattering and interference distort the incident wavefront so strongly that the spatial coherence is immensely reduced~\cite{Goodman1968book}.
The invention of optical wavefront shaping (WFS)~\cite{vel2007ol}, where $N$ multiple waves are incident on a complex sample with adjustable phases and amplitudes, has revolutionized the study of scattering of light in Nanophotonics and led to exciting applications, such as transmission optimization~\cite{vel2008prl,Popoff2010prl,aulbach2011prl,Yilmaz2019prl,Li2021pra}, light focusing~\cite{katz2011np,McCabe2011ncomms,mosk2012np,Vel2015osa,Horstmeyer2015np,Mounaix2016pra,Kubby2019Book}, light absorption and energy density control~\cite{Wan2011sci,Sarma2016prl,Ojambati2016pra,rott2017rmp}, and new biomedical imaging techniques~\cite{Yaqoob2008,Sung2009oe,Yilmaz2015optica,Kubby2019Book}. 

In the absence of absorption, the power extinguished from an incident beam is equal to the total scattered power, a well-known conservation law called the optical theorem~\cite{newt1979ajp}. 
The standard formulation of the optical theorem considers only a \textit{single} ($N = 1$) incident wave~\cite{newton1982book}. 
Naively using the single-beam optical theorem in the case of scattering with $N$ multiple incident waves, a situation typical of WFS, leads to a violation of energy conservation. 
We have recently derived a generalized optical theorem to describe the scattering and extinction by \textit{multiple} incident waves~\cite{Lagendijk20epl}. 
A crucial part in the derivation of the generalized optical theorem was the exciting discovery that multiple incident waves show cross-extinction. 
This phenomenon does not exist in common single-beam forward scattering or self-extinction, since the phenomenon is caused by interference between the scattered part of one incident beam and the coherent part of another beam. This cross-interference is always present, whether the samples are scattering or absorbing, and depending on the phases between the beams, can be constructive or destructive, making it relevant for an \textit{ab initio} description of WFS. 
 
The mutual extinction and transparency effects allow us to control the total extinction by manipulating the phase difference between the two incident beams. 
Depending on their relative angle, the extinction is varied by as much as $\pm 100\%$. 

In this paper, we present an experimental observation of the mutual extinction and transparency of two beams crossing in a scattering medium that, to the best of our knowledge, has never been observed before. 
We study the situation with $N = 2$ beams since it is the simplest form of $N$-beam interference, as is typical of wavefront shaping. 
In our experiments, we study two different kinds of samples. 
The first type of sample is a human hair, which is a biological sample with a naturally near-cylindrical cross-section~\cite{KharinSPIE2009}. 
The second type of sample is a silicon bar with a rectangular cross-section made from a crystalline-Si wafer as used in CMOS fabrication, see, \textit{e.g.}, Ref.~\cite{Prasad2019acs,Grishina2019acs}.

We compare our experimental results with an analytical approximation based on the mutual extinction theory, and we discuss when this approximation fails. 
Finally, we discuss several applications of the mutual extinction and transparency effects.

%
\section{Power flux and mutual extinction}\label{sec:currNorm}
%

\begin{figure}[tbp]
    \centering
    \includegraphics[width=0.98\columnwidth]{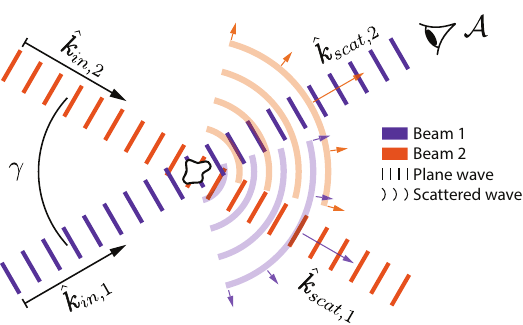}
    \caption{Schematic of two-wave mutual scattering: 
    $N=2$ beams with wavevectors $\hat{k}_{{\rm in},1}$ and $\hat{k}_{{\rm in},2}$ and mutual angle $\gamma$ are incident onto a scattering sample. 
    The scattered waves are shown as curved wavefronts to emphasize that they are present at all outgoing directions (with wavevectors $\hat{k}_{{\rm scat},1}$ and $\hat{k}_{{\rm scat},2}$), where they interfere with coherent beams leading to mutual extinction or mutual transparency. 
    The arrow colors distinguish the scattered waves and do not represent different wavelengths. 
    The detector has an area $\cal{A}$ and is placed in the far field, hence it is clear that the dimensions are not to scale. }
    \label{fig:currComp}
\end{figure}

In a light scattering experiment, a detector with area $\cal{A}$ that is placed in the direction of the wavevector $\hat{k}_{{\rm det}}$ detects in far-field the power flux or Poynting vector~\cite{Jackson1998book}. 
In the scalar-wave approximation, the flux equals the current density\footnote{For a scalar wave $\psi$, the flux $F$ is equal to $F \equiv \int_{\cal{A}} {\bm J} \rm{d}A \equiv - \int_{\cal{A}} {\rm Re}(\partial_t \psi)^*{\bm \nabla} \psi \rm{d}A$} 
${\bm J}$ integrated over $\cal{A}$ (see Fig.~\ref{fig:currComp}). 
When only one beam (beam 1) is incident in the scattering media with wave vector $\hat{k}_{{\rm in},1} = \hat{k}_{{\rm det}}$, the power flux $F$ observed by the detector is equal to~\cite{ishimaru1978book}
%
\begin{eqnarray}\label{eq:waveSplit2}
    F_{1,1} = F^{{\rm in}}_{1} - F^{{\rm ext}}_{1} + F^{{\rm scat}}_{1} ,
\end{eqnarray}
%
where $F^{{\rm ext}}_{1}$ is the flux removed from the incident flux $F^{{\rm in}}_{1}$ due to interference between the outgoing coherent wave and the scattered waves. 
Since scattered light from beam 1 is present at all angles, a fraction of the scattered flux 
$F^{{\rm scat}}_{1}$ with wavevector $\hat{k}_{{\rm scat},1} = \hat{k}_{{\rm det}}$ is also scattered into detection area $\cal{A}$. 

If a second beam (beam 2) is also incident, with a different wavevector $\hat{k}_{{\rm scat},2} \neq \hat{k}_{{\rm det}}$, \textit{i.e}, not in the direction of the detector, the power flux at the detector becomes
%
\begin{eqnarray}\label{eq:waveSplit3}
    F_{1,2} = F_{1,1} + F^{{\rm scat}}_{2} + F^{{\rm scat}}_{1,2} + F^{{\rm cross}}_{1,2}.
\end{eqnarray}
%
Here, $F^{{\rm scat}}_{2}$ is the flux of the scattered fraction of beam 2 incident into the detector with wavevector $\hat{k}_{{\rm scat},2} = \hat{k}_{{\rm det}}$, (similar to $F^{{\rm scat}}_{1}$ above). 
$F^{{\rm scat}}_{1,2}$ is the cross term describing interference between the scattered waves from both incident beams, and $F^{{\rm cross}}_{1,2}$ describes interference between the coherent wave of the incident beam 1 and the scattered wave of the incident beam 2. 
This final term of Eq.~\ref{eq:waveSplit3} corresponds to either mutual extinction or mutual transparency, depending on its sign. 
In the case of destructive interference, $F^{{\rm cross}}_{1,2}$ is negative and the total extinction is increased, corresponding to mutual extinction. 
In the case of constructive interference, the total extinction is decreased, corresponding to mutual transparency. 
This term is present for both scattering and absorbing samples. 

\begin{figure}[tbp]
    \centering
    \includegraphics[width=0.98\columnwidth]{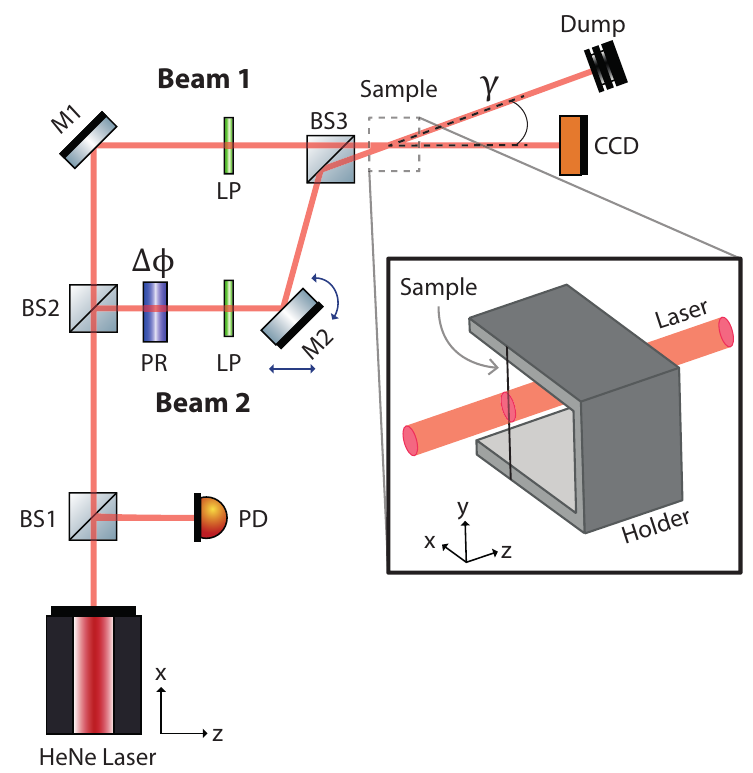}
    \caption{Schematic of the experiment. 
    The angle $\gamma$ between the two beams is controlled by moving and rotating mirror M2 and an LC phase retarder controls the phase difference $\Delta \phi$ between the two beams. 
    The inset shows the position of the sample related to the direction of beam 1. 
    (M: Mirror; PD: Photodiode; BS: Beamsplitter; LP: Linear polarizer; PR: Phase retarder.)}
    \label{fig:Setup}
\end{figure}

We experimentally obtain $F^{{\rm scat}}_{2}$ if we collect data when only the incident beam 2 is present, and we obtain $F^{{\rm in}}_{1}$ if we collect data without the scattering medium. 
Combining the data of the different situations, we get the desired interference term $F^{{\rm cross}}_{1,2}$ from the following observables 
%
\begin{eqnarray}\label{eq:crossTerms}
    F^{{\rm cross}}_{1,2}  =    F_{1,2} -  F_{1, 1}  - F^{{\rm scat}}_{2}. 
\end{eqnarray}
%
Here, we assume $F^{{\rm scat}}_{1} \ll F^{{\rm in}}_{1}$ and thus also $F^{{\rm scat}}_{1} \approx 0,\ F^{{\rm scat}}_{1,2} \approx 0$, which is reasonable because the measurement direction is equal to the incident direction ($\hat{k}_{{\rm in},1} = \hat{k}_{{\rm det}}$), and since generally coherent beams are much brighter than scattered beams~\cite{VanRossum1999rmp}. 
Using this reasonable assumption, we extract the self-extinction $F^{{\rm ext}}_{1}$ in a similar way 
%
\begin{eqnarray}\label{eq:extBeam1}
    F^{{\rm ext}}_{1}  =    F_{1,1} -    F^{{\rm in}}_{1}.
\end{eqnarray}
%
We use $F^{{\rm ext}}_{1}$ for normalization, as we want to know how the total extinction changes due to these interferences with respect to the case when only self-extinction is considered. 
Thus, we obtain the normalized total extinction 
%
\begin{eqnarray}\label{eq:totalExtinction}
    F^{{\rm TE}}  =    \frac{F^{{\rm cross}}_{1,2}}{F^{{\rm ext}}_{1}}.
\end{eqnarray}
%
It is this observable $F^{{\rm TE}}$ that reveals the desired mutual extinction and mutual transparency effects. 

\begin{figure}[tbp]
    \centering
    \includegraphics[width=0.98\columnwidth]{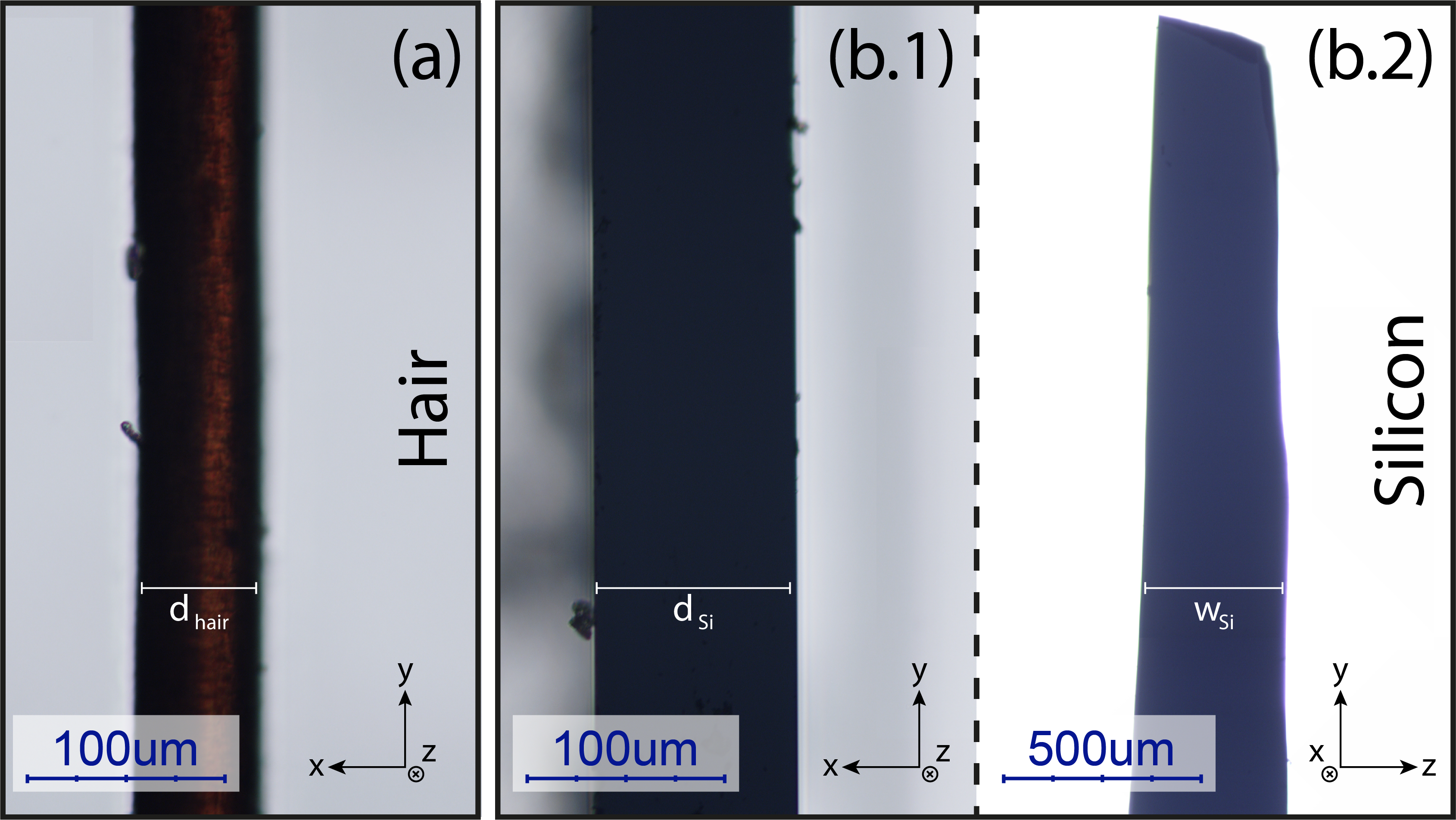}
    \caption{Microscope picture of samples used. 
    (a) picture of a human hair, from which we extract a diameter $d_{\rm{hair}} = 0.062 \pm 0.002\ {\rm mm}$. 
    (b) is separated in a picture from the side (b.1) and from the top (b.2) of the silicon bar. 
    From such microscopy pictures we extract a thickness $d_{\rm{Si}}=0.103 \pm 0.003\ {\rm mm}$ and a width $w_{\rm{Si}}=0.440 \pm 0.002\ {\rm mm}$. 
    Only one sample of each kind was used, thus one human hair and one silicon bar.}
    \label{fig:Samples}
\end{figure}

%
\section{Experimental methods}
%
To detect mutual extinction and transparency, we built the experimental setup shown in Fig.~\ref{fig:Setup}. 
A He-Ne laser ($\lambda=632.8\ \rm{nm}$) is used as a source. 
The laser beam is split into two incoming beams, beam 1 and beam 2, in a modified Mach-Zehnder configuration with a slight and controllable skewness at the outgoing beamsplitter  BS3. 
Before reaching BS3, both beams pass through linear polarizers (LPVIS050 Thorlabs), and beam 2 passes through a liquid crystal phase retarder (LCC1413-B Thorlabs), which we use to control the phase difference $\Delta \phi$. 
By carefully moving and rotating mirror M2, we control the angle $\gamma$ between the two beams. 
At a fixed location downstream of BS3, where the sample is located, the two beams cross at an angle $\gamma$.
We use a CCD camera to detect beam 1 by integrating over the illuminated pixels, and a photodiode at the beginning of the optical circuit to correct for laser fluctuations. 
Both the angle variation and the phase variations are made in beam 2, whereas only beam 1 is detected with the CCD camera. 
At every angle $\gamma$, the phase $\Delta \phi$ was varied from $0$ to $2\pi$ and back. 
For each phase, we took three consecutive exposures with the CCD camera to average over environmental noise.

\begin{figure}[tbp]
    \centering
    \includegraphics[width=0.98\columnwidth]{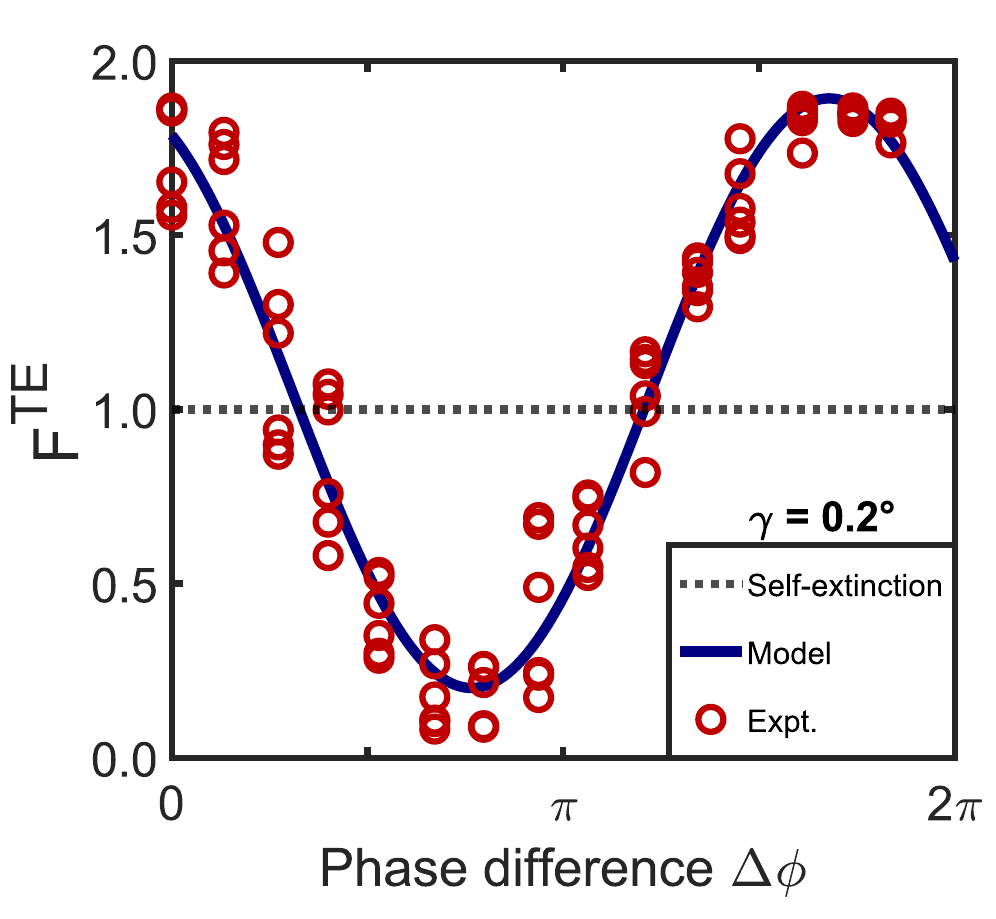}
    \caption{Total extinction $F^{{\rm TE}}$ versus phase difference $\Delta \phi$ for an angle $\gamma=0.20^{o}$ between the two beams for the human hair. 
    Dark red circles are experimental data while the solid blue line is a fit to the exact mutual extinction model. 
    Dotted line is the effect in case of no mutual extinction or transparency. 
    The phase difference is extracted from the retardance of the phase retarder.}
    \label{fig:phaseHair01}
\end{figure}

We position the sample in the intersection plane of the two beams, as shown in the inset of Fig.~\ref{fig:Setup}. 
All samples we study have a bar-like geometry, meaning that one dimension is much larger than the other two. 
We define the active area as the overlapping area of the two beams in the illuminated face of the sample. 
Thus, in the \textit{x}-direction the dimension of the active area is given by the geometry of the sample, and in the \textit{y}-direction the dimension is given by the beam diameter.

We study two different samples at present: 
The first sample is a human hair, which has a natural near-cylindrical shape with a diameter of $d_{\rm{hair}} = 0.062 \pm 0.002\ {\rm mm}$ as observed with a microscope (see Fig.~\ref{fig:Samples}(a)). 
The scattering properties of human hair are of special interest for the 3D animation industry to obtain a realistic simulation of hair in animated characters~\cite{Marschner2003acm}. 
Furthermore, single human hair fibers are widely used to teach light diffraction in undergraduate and secondary education, approximating it to the inverse of a single slit~\cite{Messer2018tpt}. 
The second sample is a thin silicon bar cleaved from a standard Complementary Metal Oxide Semiconductor (CMOS) wafer, with a thickness $d_{\rm{Si}} = 0.103 \pm 0.003\ {\rm mm}$ and a width $w_{\rm{Si}} = 0.440 \pm 0.002\ {\rm mm}$ (see Fig.~\ref{fig:Samples}(b)). 
The scattering properties of silicon are highly relevant for the semiconductor industry since it is the main material used in electronics. 
To limit the complexity we use samples that are translational invariant along the y-axis. 
To accomplish this, the silicon bar is illuminated from the side and not from the top.

%
\section{Experimental results}
%
%

First, let us consider the case where the angle $\gamma$ is fixed. 
In Fig.~\ref{fig:phaseHair01}, we show the normalized total extinction of a human hair while changing the phase $\phi$. 
The angle between incoming beams is $\gamma = 0.2^o$. 
We see that the extinction follows a cosine-like trend, as expected from the prediction~\cite{Lagendijk20epl} (See Appendix~\ref{sec:app}). 
We also see that at the maximum the extinction nearly doubles, while for the minimum the extinction is close to zero, so that the object appears nearly fully transparent. 
In theory, the total extinction is minimum for $\Delta\phi = 0$ and maximum for $\Delta\phi = \pi$. 
We see that the experimental data are shifted in phase, which arises from uncertainty in the true-zero phase (see Appendix~\ref{sec:app2}). 
This effect is due to changes in the optical path of beam 2 when moving and rotating mirror M2 (see Fig.~\ref{fig:Setup}). 
Still, we see that the periodicity of the observed cosine-like curve agrees well with theory, and thus this phase offset does not affect the final results. 

\begin{figure}[tbp]
    \centering
    \includegraphics[width=0.98\columnwidth]{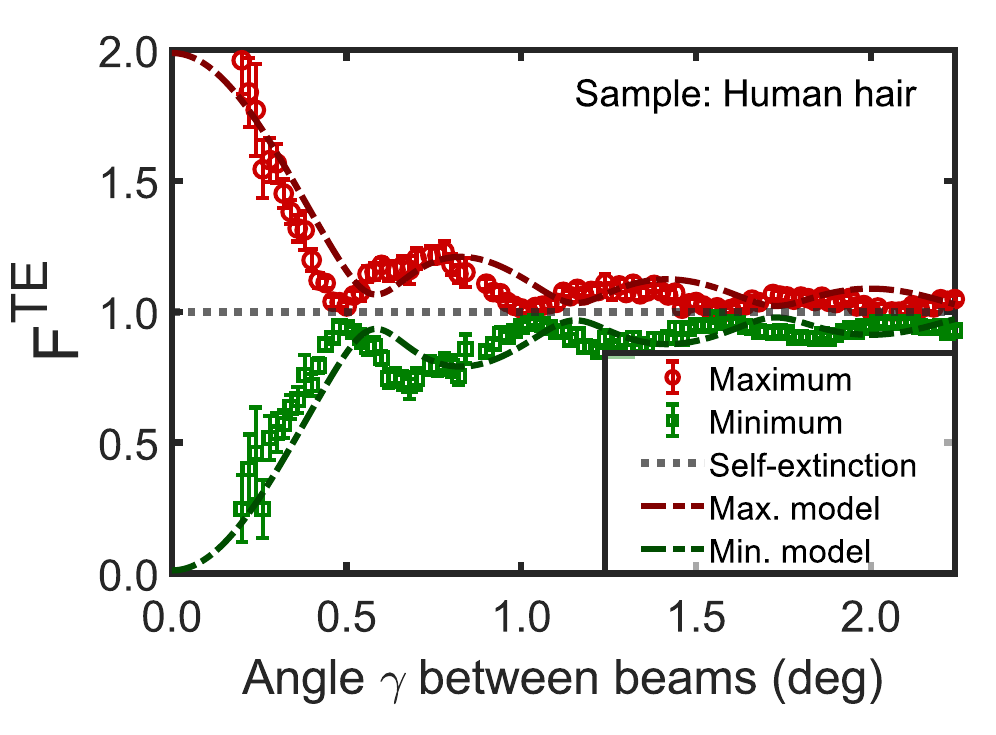}
    \caption{Total extinction versus angle $\gamma$ between beams for a human hair. 
    The red circles (green squares) correspond to the maximum (minimum) extinction obtained from phase variations (see Fig.~\ref{fig:phaseHair01}). 
    The dashed lines are our analytical model with no adjustable parameter (see Appendix~\ref{sec:app}).}
    \label{fig:angleHair01}
\end{figure}

In Fig.~\ref{fig:angleHair01}, we plot the total extinction of the human hair versus angle $\gamma$. 
The symbols correspond to the maximum and minimum extinctions obtained as a function of the phase at each angle (see Fig.~\ref{fig:phaseHair01}). 
The observed maximum and minimum extinctions show an oscillatory behavior versus incident angle $\gamma$, typical of interference between scattered and coherent beams. 
Along with the experimental data, our analytical model is shown to be in good agreement. 
This model uses the scattering amplitude of an impenetrable flat surface (see Appendix~\ref{sec:app}) as an approximation to obtain the variations in the forward scattering due to the mutual extinction effect. 
Our model has no adjustable parameter since the width of the sample $a_{mic}$ is obtained by microscopy inspection.
We see that the data follows a sinc trend similar to the model, with a slight shift in angle discussed in the next section.

\begin{figure}[tbp]
    \centering
    \includegraphics[width=0.98\columnwidth]{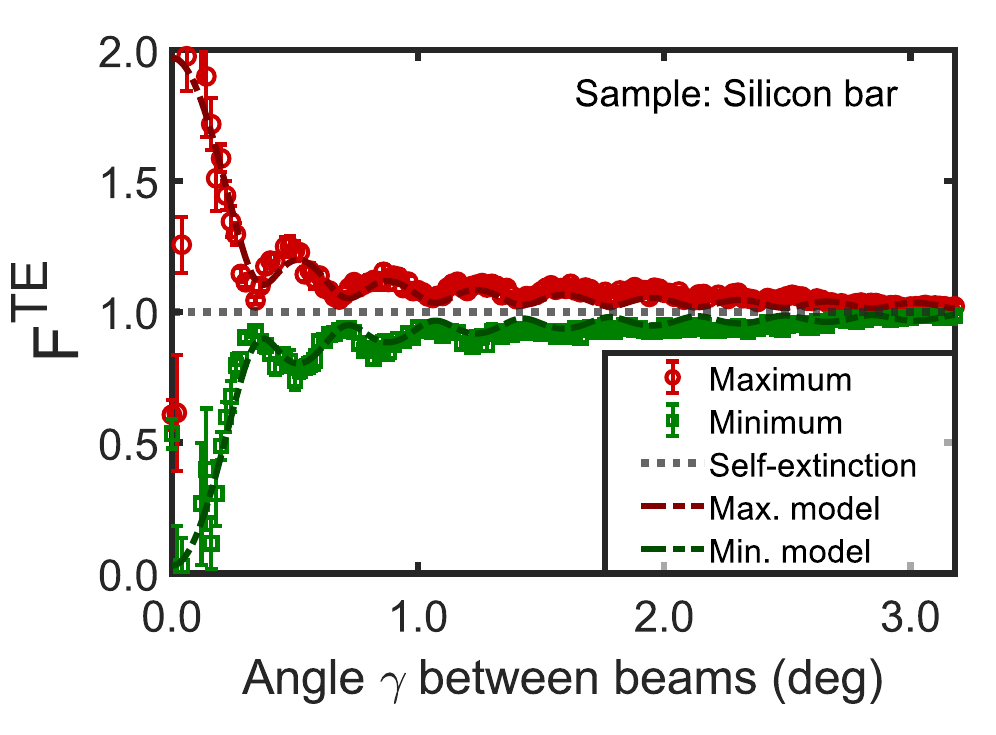}
    \caption{Total extinction versus angle $\gamma$ between beams for a silicon bar. 
    The red circles (green squares) correspond to the maximum (minimum) extinction obtained from phase variations. 
    The dashed lines are our analytical model with no adjustable parameter.}
    \label{fig:angleSilicon01}
\end{figure}

For the silicon bar, the total extinction against angle is shown in Fig.~\ref{fig:angleSilicon01}.
The data has a similar sinc shape as the human hair, but here the frequency of the interference fringes is higher. 
Furthermore, the analytical approximation agrees even better with the experimental data than for the hair sample.

%
\section{Discussion}
%

%
\subsection{Observations and model}
%
In both the phase and angle variation experiments, we obtained the trend predicted by the mutual extinction theory~\cite{Lagendijk20epl}, namely a cosine-like trend in the phase variation experiments and a sinc-like trend in the angle variation experiments. 
We see that using our analytical model, we obtain a description of the mutual extinction, which gives a faithful interpretation of the observations. 

In Fig.~\ref{fig:phaseHair01}, we see that the modulation of the total extinction is close to the full range from 0 to 2. 
To discern how large this modulation is compared with the total signal of the beam, we obtained that in the case of the human hair, the extinguished light amounts to about 15.4\% of the incoming light. 
With the mutual extinction and transparency effect, the extinguished light varies from approximately 2.5\% to 24\%. 
This is close to the maximum modulation predicted by theory, which varies from 0\% to 30.8\%. 

In Fig.~\ref{fig:angleHair01}, we observe that the model predicts a slightly lower frequency of fringes than measured, meaning the nodes of the model are located at larger angles than the ones from the experimental data. 
The curve shapes are in excellent overall agreement. 
The small deviations emerge from the fact that the human hair has a cylindrical geometry. 
Thus, the sample thickness varies with the lateral position within the incident beams, which is not addressed in the model, where we assumed a flat impenetrable surface. 
We quantify this deviation using the width of the sample $a$ as a single adjustable parameter in our model (see Eq.~\ref{eq:appModel}), and compare the estimated value $a_{\rm mod}$ with the width of the sample used originally as input for the model, which we extracted from optical microscopy inspection $a_{\rm mic}$. 
In Table~\ref{tab:widths} we see from this comparison that the adjusted width does not match with the independent observation. 

\begin{table}[ht]
\centering
\begin{tabular}{p{0.24\columnwidth} | p{0.33\columnwidth} | p{0.33\columnwidth}}
\textbf{Sample} & \textbf{$a_{\rm mod}$ ({\rm {\rm mm}})} & \textbf{$a_{\rm mic}$ ({\rm mm})} \\ \hline
Hair            & $0.072 \pm 0.001$                 & $0.062 \pm 0.002$                \\
Silicon         & $0.105 \pm 0.001$                & $0.103 \pm 0.003$                
\end{tabular}
\caption{Table of sample dimension extracted from the model ($a_{\rm mod}$), and extracted from optical microscopy inspection ($a_{\rm mic}$). Error range of the model is due to the dispersion of the experimental data, while the error range of the optical microscopy inspection is due to microscope resolution.}
\label{tab:widths}
\end{table}

\begin{figure}[tbp]
    \centering
    \includegraphics[width=0.98\columnwidth]{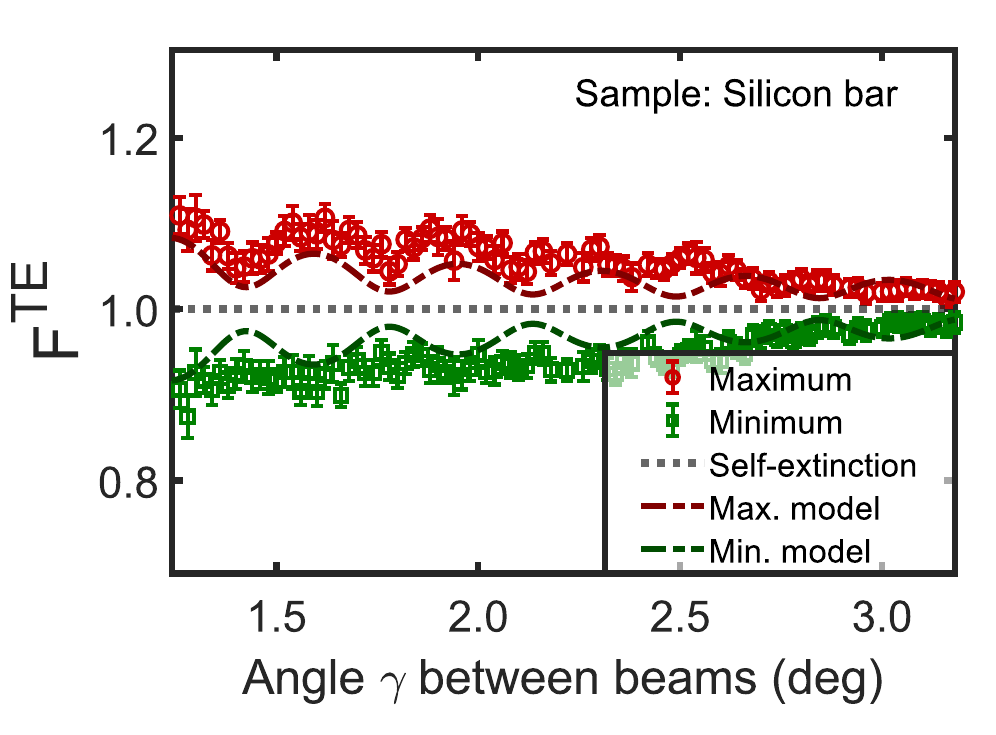}
    \caption{Total extinction versus angle $\gamma$ between beams for a silicon bar. 
    Zoom in for large angles from Fig.~\ref{fig:angleSilicon01}. 
    The dashed lines are our analytical model with no adjustable parameter.}
    \label{fig:angleSilicon03}
\end{figure}

In contrast, for the silicon sample, which has a box-like geometry, $a_{\rm mod}$ and $a_{\rm mic}$ are equal within the error bars. 
We see in Fig.~\ref{fig:angleSilicon01} that for small $\gamma$ the analytical model describes the silicon data accurately. 
This can lead one to think that this approximation describes the results completely, but this is not the case. 
In Fig.~\ref{fig:angleSilicon03}, we zoom in on the silicon results to observe in detail the deviation for angles between $\gamma = 2$ to $3^o$. 
We see that for $\gamma > 2.0^o$ the trend followed by the experimental data has some discrepancies with the model: 
In the beginning, the nodes start to shift, and finally, the shape of the sinc curve is completely lost. 
Importantly, at all these angles the measured phase dependencies keep following a cosine-like curve as in Fig.~\ref{fig:phaseHair01}, and the signals are significantly larger than the errors. 
We extended our analytical model by also taking into account the beam divergence of the laser using a convolution over the Gaussian beam profile. Nevertheless, we see that the average caused by divergence does not explain this deviation. 

The apparent random shape of the total extinction at larger angles is probably an indication that we are entering the speckle regime where the variations in the extinction do not depend only on the dimensions of the sample, but also on the detailed spatial distribution of the scatterers inside the material or the surface roughness. 
In the speckle regime, the depth of the sample also plays an important role. 
For the case of the silicon bar, when changing the angle, the path length inside the sample also changes, which modifies the scattering amplitude in that direction. 
This is not accounted for in the model, where we assume a flat impenetrable surface.

When the sample is fully opaque and its size is much larger than the wavelength, the {\it angular dependence} of the scattering amplitudes, and thus of the mutual extinction and transparency, will follow the conventional dependence of the Kirchhoff integral~\cite{Born1970book}, which we used to derive our model. 
In the situation where the sample is not fully opaque, as in our experiments, the angular dependence of the scattering amplitudes will become speckle-like, starting at the large angles of $\gamma$. 
From theory~\cite{Lagendijk20epl}, we expect the speckle regime to extend for very large angles, including for angles larger than $90^{o}$. 

If the sample has a mean free path $l$ much larger than its size ($l \gg L$), the angular speckle will set out in the immediate vicinity of the first zero-crossing, which itself is determined by the geometry of the sample. 
The more transparent the sample is, the more prominent the speckle will be. 
But when averaging over many realizations, the result of the Kirchhoff integral is recovered.
%
\subsection{Applications}
%
The mutual extinction and transparency effect was discovered when explaining the violation of energy conservation when simulating WFS. 
For that reason, an important application of mutual extinction and transparency is in WFS. 
We think mutual extinction and transparency can help on the discussion on open channels in complex media~\cite{Sarma2016prl} and in the simultaneous optimization of transmitted and reflected intensity, both currently studied with WFS. 
Furthermore, we think wavefront modulation can be used to design a non-diffracting beam shape, such as Bessel beams~\cite{Cizmar2009oe,Gong2013ao}, which can be used in mutual extinction and transparency in extended samples. 

The mutual extinction and transparency effect can have applications in the field of ultraviolet communications for non-line-of-sight links~\cite{Raptis16osa}, which is based on light scattering in the atmosphere. 
The scattered light can be enhanced or reduced if the transmitter uses two beams that cross in the active area. 
Furthermore, mutual transparency can be used to reduce losses related to light attenuation in the atmosphere. 
Deeper research is needed to translate the results presented in this paper to a more practical system. 

The mutual extinction and transparency effect is a promising tool to use the observed interference fringes to infer the shape and size of an object, including free-form samples~\cite{Haberko2013pra}.

Diffusion Wave Spectroscopy (DWS) has become a popular technique for studying time-dependent optical properties of complex materials and for bio-imaging~\cite{maret1987zpcm,pine1988prl,Stetefeld2016br}. 
Unfortunately, DWS is not well suited for samples that absorb rather than scatter. 
To study the motion of scattering particles, one currently uses techniques such as Dynamic Light Scattering (DLS)~\cite{Berne2013book}, which is also not suited for samples that absorb rather than scatter. 
In contrast, the mutual extinction and transparency effect can be used for samples that scatter or absorb (or both), as long as there is some detectable intensity left of the incoming beam. 
We envisage for those samples that dynamics can be probed by using a time-dependent mutual extinction technique.

%
\section{Conclusions}
%

We measured the total extinction of two beams crossing through a scattering object, namely a human hair and a silicon bar. 
Upon varying the relative angle and phase between the beams, we measured the variations in the total extinction. 
When the angle is close to zero, we control the extinction in such a way that the scattering object is almost twice as opaque or nearly fully transparent to the beams. 
Alternatively, if the angle is larger we enter, for non-opaque samples, into the speckle regime where fluctuations of the mutual extinction depend on the precise shape and distribution of the scatterers distribution in the sample.

Our results are in close agreement with the recently presented mutual extinction theory~\cite{Lagendijk20epl}, turning this experiment into the confirmation of this effect. 
We used an analytical approximation of the scattering amplitude applicable when the sample is opaque. We have seen that this approximation is a good model for box-like geometries and small angles, but at the same time, we see that the mutual extinction effect cannot faithfully be interpreted with such a simple approximation of the scattering amplitude, for transparent samples.

%
\begin{acknowledgments}
%

We thank Cornelis Harteveld for expert technical support and sample preparation, Manashee Adhikary and Matthijs Velsink for helpful discussions, and Catalina Garc\'ia for her contribution. 
This work was supported by the NWO-TTW program P15-36 "Free-form scattering optics" (FFSO) and the MESA+ Institute section Applied Nanophotonics (ANP).
%
\end{acknowledgments}
%
%
\appendix
\section{Model for angular dependence}\label{sec:app}
%

In Ref.~\cite{Lagendijk20epl}, we have calculated and described the scattering amplitudes of both a collection of point dipoles and a flat and opaque object. 
Here, we compare our experimental results with the latter, where the scattering amplitude is derived as (see Eq.~8 of Ref.~\cite{Lagendijk20epl}) 
%
\begin{eqnarray}\label{eq:appFDF}
    f  = \frac{iab}{\lambda} \rm{sinc}(\alpha) \, \rm{sinc}(\beta)
\end{eqnarray}
%
where $\lambda$ is the wavelength in vacuum, $\alpha \equiv \frac{a}{2} (  \cos \theta_{x, {\rm out} }    -  \cos \theta_{x, {\rm in} }  )$,
$\beta \equiv \frac{ b}{2 } ( \cos \theta_{y, {\rm out} } - \cos \theta_{y, {\rm in} } )$, $\theta_{x,{\rm in}}$ and $\theta_{y,{\rm in}}$ are the angles of the incident waves with respect to the \textit{x} and \textit{y}-axes, $\theta_{x,{\rm out}}$ and $\theta_{y,{\rm out}}$ are the angles of the outgoing waves with respect to the \textit{x} and \textit{y}-axes, and $a$ and $b$ are the dimensions of the sample in the \textit{x} and \textit{y}-directions, respectively.

The scattering amplitude $f$ is simplified if we consider both incident waves to be in the \textit{xz}-plane. 
Consequently, 
 the power flux $F$ is simplified to
%
\begin{eqnarray}\label{eq:appModel}
    F_{\rm{model}}  =  \textrm{sinc} \left( \frac{2 \pi a}{\lambda} \cdot \textrm{sin} \frac{\gamma}{2}\right),
\end{eqnarray}
%
For the human hair, the dimension $a$ is defined as the diameter of the hair ($a=d_{\rm{hair}}$), and for the silicon bar it is defined as the thickness of the bar ($a=d_{\rm{Si}}$). 

When the sample is absorbing, as in the case of our experiments, the amplitude of $F$ is strongly affected by the absorption coefficient of the sample. 
This is not accounted for in Eq.~\ref{eq:appModel}, where the amplitude is 1. 
Nevertheless, when normalizing with the self-extinction, as it is done in Eq.~\ref{eq:totalExtinction}, this effect is canceled out. 

To take the divergence $\Theta$ of the laser beam into account, we implemented a convolution between $F_{\rm{model}}$ and the angular momentum profile of a Gaussian beam $U(\theta)$ given by~\cite{Nieminen2008jpa}
%
\begin{eqnarray}\label{eq:Gprofile}
    U(\theta) = \rm{exp}\left ( - \rm{tan}^2(\theta) / \rm{tan}^2(\Theta) \right ).
\end{eqnarray}
%
$F_{model}$ includes both positive and negative values, meaning both mutual extinction and mutual transparency. 
To separate between both cases, we take $F_{max} = 1+|F_{model}|$ as the maximum curve and $F_{min} = 1-|F_{model}|$ as the minimum curve, which correspond, respectively, to the dashed red curve and the dashed green curve in Figures~\ref{fig:angleHair01},~\ref{fig:angleSilicon01}, and~\ref{fig:angleSilicon03}. 

%
\section{Model for phase dependence}\label{sec:app2}
%
When the angle $\gamma$ is fixed, if we change the relative phase $\Delta \phi$, the mutual extinction effect fluctuates following a cosine function. 
If $\gamma$ is outside the speckle regime, the total extinction is minimum for $\Delta\phi = 0$ and maximum for $\Delta\phi = \pi$. 
Differently, if $\gamma$ is inside the speckle regime, deviations in the position of maximum total extinction are a reflection of the complex part of the scattering amplitude $f$. 

In our current experiment, the path length of beam 2 changes when we change the angle, meaning is not possible to retrieve the true phase for the maximum total extinction. 
Instead, to compare our experimental results with the model we use
%
\begin{eqnarray}\label{eq:cosFit}
    F^{{\rm TE}}_{\rm fit}  =    1+c_1 \cdot \cos(c_2\Delta\phi - c_3\pi) + c_4,
\end{eqnarray}
%
where $c_1=0.8461$, $c_2=1.0954$, $c_3=0.7800$, and $c_4=0.0464$ are adjustable parameters. 
We can extract from here $\Delta \phi_{\rm off} = 0.78\pi$ as the phase offset in the measurements. 
%
%

\bibliographystyle{apsrev4-2}
\bibliography{extinction.bib}

\end{document}